\documentclass[12pt]{article}
\usepackage{jheppub} 

\pdfoutput=1
\usepackage{amsmath,amssymb,amsfonts}
\usepackage{graphicx}
\usepackage{braket}
\usepackage{booktabs}
\usepackage[]{hyperref}
\usepackage[sort&compress]{natbib}
\usepackage{tikz}
\usepackage{cleveref}
\usepackage{totcount}
\newtotcounter{citnum} 
\def\oldbibitem{} \let\oldbibitem=\bibitem
\def\bibitem{\stepcounter{citnum}\oldbibitem}
\newcounter{qnumber}

\newcommand{\bk}[1]{\left<#1\right>}

\renewcommand{\[}{\begin{equation}\begin{aligned}}
\renewcommand{\]}{\end{aligned}\end{equation}}
\DeclareMathOperator{\HeunC}{HeunC}

\begin{document}

\title{The Radial Action from Probe Amplitudes to All Orders}
\author[a]{Uri Kol,}
\emailAdd{urikol@gmail.com}
\affiliation[a]{Center for Cosmology and Particle Physics, Department of Physics
New York University, 726 Broadway, New York, NY 10003, USA}
\author[b]{Donal O'Connell,}
\emailAdd{donal@ed.ac.uk}
\affiliation[b]{Higgs Centre for Theoretical Physics, School of Physics and Astronomy, Peter Guthrie Tate Road, The University of Edinburgh, EH9 3FD, Scotland}
\author[c,d]{ and Ofri Telem}
\emailAdd{t10ofrit@gmail.com}
\affiliation[c]{Department of Physics, University of California, Berkeley, CA 94720, USA}
\affiliation[d]{Ernest Orlando Lawrence Berkeley National Laboratory, Berkeley, CA 94720, USA}

\abstract{
We extract the relativistic classical radial action from scattering amplitudes, to all orders in perturbation theory, in the probe limit. Our sources include point charges and  monopoles, as well as the Schwarzschild and pure-NUT gravitational backgrounds. A characteristic relativistic effect, that scattering trajectories may wind around these sources any number of times, can be recovered when all-order amplitudes are available. We show that the amplitude for scattering a probe off a pure NUT is given by the solution of a transcendental equation involving continued fractions, and explain how to solve this equation to any desired loop order. 
}

\maketitle
\section{Introduction}
Recent years have seen tremendous progress in the field of Post-Newtonian (PN) and Post-Minkowskian (PM) Effective Field Theory (EFT), relevant for the analytic calculation of compact object coalescence events.
For comprehensive reviews of PN methods in GR, see \cite{Blanchet:2004ek,Porto:2016pyg,Levi:2018nxp} and references within, as well as more recent progress up to 6-7PN \cite{Bini:2020hmy,Bini:2020nsb,Bini:2020rzn,Bini:2020uiq,Bini:2020wpo,Bini:2021gat}. This progress has been of fundamental importance for reliable analytic and semi-analytic determination of gravitational waveforms, especially in the context of Effective-One-Body Hamiltonians \cite{Buonanno:1998gg,Buonanno:2000ef,Damour:2001tu,Damour:2008qf,Arun:2008kb,Damour:2009kr,Damour:2009wj,Pan:2010hz,Taracchini:2013rva,Bini:2014zxa,Nagar:2019wrt,Antonelli:2019fmq,Nagar:2019wds,Albanesi:2021rby}.

Besides the more conventional PN methods, there has also been a renewed interest in PM methods \cite{Bjerrum-Bohr:2013bxa,Bjerrum-Bohr:2014zsa,Vines:2017hyw,Bjerrum-Bohr:2018xdl,Cheung:2018wkq,Vines:2018gqi,Damgaard:2019lfh,Cristofoli:2019neg,Bjerrum-Bohr:2019nws,Antonelli:2019ytb,Brandhuber:2019qpg,Guevara:2019fsj,Maybee:2019jus,Arkani-Hamed:2019ymq,DiVecchia:2019myk,Bjerrum-Bohr:2019kec,DiVecchia:2019kta,Chung:2019duq,Damour:2020tta,Cristofoli:2020uzm,Kalin:2020fhe,Cheung:2020gyp,Kalin:2020mvi,DiVecchia:2020ymx,delaCruz:2020bbn,AccettulliHuber:2020dal,Guevara:2020xjx,DiVecchia:2021ndb,Bautista:2021wfy,Damgaard:2021rnk,Bjerrum-Bohr:2021din,Bjerrum-Bohr:2021vuf,DiVecchia:2021bdo,Liu:2021zxr,Brandhuber:2021kpo,Dlapa:2021npj,Cristofoli:2021vyo,Brandhuber:2021eyq}, which involve an expansion in Newton's constant $G$ but not in $v/c$. A central idea which underlies much of the PM approach (as well as some of the PN approach) is the matching of General Relativity (GR) to an effective field theory \cite{Goldberger:2004jt,Goldberger:2005cd}. This matching can be done at the level of classical observables such as the scattering angle \cite{Damour:2017zjx,Damour:2019lcq} or the impulse \cite{Guevara:2017csg,Kosower:2018adc}.

One promising approach to PM dynamics involves an unexpected connection --- to relativistic scattering amplitudes in quantum field theory~\cite{Donoghue:1993eb,Donoghue:1994dn,Neill:2013wsa,Bjerrum-Bohr:2013bxa,Cachazo:2017jef,Guevara:2017csg,Cheung:2018wkq,Kosower:2018adc}.
Since gravitons couple with a $\sqrt{G}$ factor, the PM expansion goes over to the usual perturbative loop expansion of the corresponding scattering amplitude, with the $(n-1)$-loop amplitude responsible for the $n$-PM EFT coefficients. 
The basic advantage of the use of scattering amplitudes is they can be evaluated using cutting edge tools such as generalized unitarity \cite{Bern:1994cg,Bern:1994zx}, the double copy~\cite{Bern:2008qj,Bern:2010ue,Bern:2010yg}, as well as sophisticated loop integration methods \cite{Smirnov:2008iw}. Impressive concrete progress in this direction has been made~\cite{Cheung:2018wkq,Bern:2019crd,Bern:2019nnu,Bern:2020buy,Bern:2020gjj,Bern:2021dqo} by matching full GR to an effective 2-body theory at the level of quantum scattering amplitudes, or more precisely their $\hbar\rightarrow 0$ limit. 
Though very promising, the on-shell approach is not without its difficulties, especially at order $G^4$ and above where it need to capture tail effects \cite{BonnorRotenberg66,BlanchetDamour86,BlanchetDamourTail,BlanchetDamour92,Blanchet_1993,Blanchet:1997jj,Asada:1997zu,Galley:2015kus,Marchand:2016vox} which are non-local in time, e.g. an outgoing gravitational wave which is reflected in the far zone back into the inspiral. 

A striking feature of the PM EFT approach has been the importance of the \textit{radial action} $I_r(r)\equiv\int_{r_{\text{turn}}}^r p_r(r^{\prime}) dr^{\prime}$, which arises in the classical Hamilton-Jacobi equation for the system \cite{landau1982mechanics}. This should not come as a surprise, as the radial action encapsulates all of the classical dynamics of the system. In \cite{Kalin:2019inp,Kalin:2019rwq}, the radial action played a central role in the Boundary-to-Bound map, which allowed for the analytic continuation of scattering data to dynamical invariants for generic
(bound) orbits. This was further utilized to extract compact binary dynamics up to 4PM order \cite{Kalin:2020fhe,Kalin:2020lmz,Kalin:2020mvi,Dlapa:2021npj}.
Inspired by the eikonal approximation, the authors of \cite{Bern:2021dqo} suggest the following amplitude-radial action relation
\begin{eqnarray}\label{eq:ampradac}
i\mathcal{A}\,\sim\,\int_{j}\left(e^{i I_{r}}-1\right)\,.
\end{eqnarray}
The aim of this paper is to derive this relation in the classical limit and in the probe limit $\kappa\equiv\frac{m_1 m_2}{(m_1+m_2)^2}\rightarrow 0$, to all orders in $G/j$. We do this by solving the QM scattering problem for a probe mass in curved space --- the same calculation which leads to BH greybody factors \cite{Unruh:1976fm,Sanchez:1977si,Sanchez:1977vz} and quasinormal modes \cite{Teukolsky:1973ha,Teukolsky:1974yv}. However, we are interested in the classical $\hbar\rightarrow 0$ limit of this calculation, which we take following the seminal work of \cite{Ford1959,Berry_1972}. This limit exposes a simple relation between the quantum phase shifts $\bar{\delta}_{\bar{j}}$, the azimuthal scattering angle $\Delta\varphi$ and the radial action, namely:
\begin{eqnarray}\label{eq:relation}
\delta_{j}-\frac{\pi j}{2}+c(\hbar)=\lim_{r\rightarrow\infty} \left[I_{r}(r)\,-\,d(r)\right]~~,~~\Delta\varphi=\pi-2\frac{d \delta_{j}}{dj}\,.
\end{eqnarray}
In these expressions, $\delta_{j}=\lim_{\hbar\rightarrow 0}\, \hbar\bar{\delta}_{\bar{j}}\,|_{\bar{j}=j/\hbar-\frac{1}{2}}$ is the saddle point value of the phase shift $\bar{\delta}_{\bar{j}}$; $c(\hbar)$ is a $j$ independent constant, which emerges from the regularization of $\delta_{j}$ in the $\hbar\rightarrow0 $ limit, while $d(r)=kr+\eta\log(2kr)$ is a universal function describing the accumulation of phase as $r\rightarrow\infty$. 

In this paper we calculate the probe-limit phase shifts $\delta_j$ explicitly by solving a relativistic wave equation in a background gauge field/metric. See~\cite{Bautista:2021wfy} for a related connection between wave equations and classical scattering. Indeed the connection between scattering amplitudes, quantum mechanics, and classical point particles has been a recurring theme in the literature. Evidently, point-particle actions of the kind taught in undergraduate classes worldwide are somehow connected to relativistic quantum electrodynamics and quantised (effective) general relativity. The relation is simply that point-particle actions emerge as EFTs in a long-distance, low-energy limit for localised particles as was emphasised by Goldberger and Rothstein~\cite{Goldberger:2004jt}. Therefore the quantum \emph{mechanics} of these worldline actions captures the relevant dynamics. In the probe limit, fully non-perturbative amplitudes are available by solving the relevant (relativistic) Schr\"odinger equation; formerly confusing issues related to pair production are nowadays well understood and need not concern us. Our work is closely connected to other approaches based on studying the quantum field theory of the worldlines~\cite{Mogull:2020sak,Jakobsen:2021smu,Shi:2021qsb}.

The outline of our paper is as follows. First, we review our setup in section~\ref{sec:setup}, and also review key quantities and their dimensions when $c=1,\,\hbar\neq1$. In section~\ref{sec:scalar} we re-derive the relation \eqref{eq:relation} in a modern language, following the work of \cite{Ford1959,Berry_1972}. We make use of this relation to derive the all order classical scattering angle for relativistic Coulomb scattering (a.k.a Darwin scattering) in section~\ref{sec:Coulo}, including non-perturbative effects. In section~\ref{sec:Schwarzschild} we follow the steps taken for Coulomb scattering, and examine how the phase shifts for the scattering of a scalar in a Schwarzschild background converges in the classical limit to the relation \eqref{eq:relation}. 

Next, we generalize \eqref{eq:relation} to the scattering of particles with spin, as well as to the scattering of monopoles and charges. The case of monopole scattering is unique in that there is an extra angular momentum carried by the electromagnetic field \cite{Thomson,Lipkin:1969ck,Boulware:1976tv,Schwinger:1976fr,Shnir:2005xx}. The non-perturbative effect of this angular momentum on monopole scattering amplitudes was recently captured in the electric-magnetic S-matrix construction of \cite{Csaki:2020inw}, based on pairwise helicity \cite{Csaki:2020uun}. In section~\ref{sec:monopole} we take the classical limit of this S-matrix for the scattering of a Coulomb charge in a monopole background, and reproduce the classical scattering angle to all orders.

A last, highly non-trivial check of our formalism is presented in section~\ref{sec:NUT}, where we calculate the all-order scattering angle for a probe mass in NUT space. This is the gravitational double copy of charge-monopole scattering \cite{Luna:2015paa,Caron-Huot:2018ape,Huang:2019cja,Alawadhi:2019urr,Kol:2020ucd,Moynihan:2020gxj,Emond:2020lwi,Kim:2020cvf,Alawadhi:2021uie}. Here we find the all-order solution of the quantum scattering problem in terms of prolate spheroidal functions, and show how its phase shifts exactly reproduce the classical scattering angle to all orders in the PM expansion. Our results allows us to draw the following conclusions:
\begin{itemize}
\item Probe mass-NUT scattering is the double copy of charge-monopole sctatering, with $\bar{q}=\hbar^{-1}2E_{\text{probe}}G\ell_{\text{NUT}}$ playing the role of pairwise helicity \cite{Csaki:2020inw}, and is quantized in half integer units.
\item The phase shift formalism captures the full non-perturbative dynamics of probe mass-NUT (charge-monopole) scattering, including the angular momentum in the gravitational (electromagnetic) field.
\item The quantum-classical correspondence is based on a non-trivial number theoretical relation between spheroidal eigenvalues and elliptic integrals.
\end{itemize}

Finally, in section~\ref{sec:outlook} we conclude our discussion and also briefly discuss future prospects for applying our formalism beyond the probe limit, as a novel quantum approach to self-force corrections \cite{Mino:1996nk,Quinn:1996am,Barack:1999wf,Barack:2001gx,Detweiler:2002mi,Barack:2007tm,Gralla:2008fg,Barack:2009ux,Damour:2009sm,Barack:2010tm,Blanchet:2010zd,Barack:2010ny,Barack:2018yly,Barack:2018yvs}. 

\section{Setup and Notation}\label{sec:setup}

Our ultimate goal is to describe the scattering of two classical particles with arbitrary classical spins. Since we work in the probe limit, we can always map the scattering problem to an equivalent one body problem, in which a probe particle moves in the presence of a central classical electric/magnetic/gravitational field. Classically, we are interested in the trajectories of this particle, and in particular in the classical scattering angle $\chi$ defined as
\begin{eqnarray}
\cos\left(\chi\right)=\hat{p}_{\text{in}}\cdot\hat{p}_{\text{out}}\,,
\end{eqnarray}
where $\hat{p}_{\text{in}}(\hat{p}_{\text{ot}})$ is the direction of the momentum of the incoming (outgoing) particle.
In the spinless (and non-magnetic) case with only orbital angular momentum, the entire effective one body motion takes place in the plane transverse to the total angular momentum $\vec{J}$. When talking about classical trajectories, it is convenient to take $\hat{z}\,\left|\right|\,\vec{J}$ and $\hat{p}_{\text{in}}=-\hat{x},\,\hat{p}_{\text{out}}=\cos(\Delta\varphi)\hat{x}+\sin(\Delta\varphi)\hat{y}$, so that $\chi=\Delta\varphi$.

In more general cases with spin, the motion does not take place in the plane, and so generally $\chi\neq\Delta\varphi$.

Since we are interested in taking the classical limit of quantum $2\rightarrow2$ scattering, it is worthwhile to describe the setup there too. The quantum problem is also trivially reduced to an effective one body problem. Unlike the classical one, here we take the incoming state as a plane wave in the $\hat{z}$ direction. In this case, the amplitude squared gives the probability to scatter to a given angle $\theta$. Taking the classical limit and we expect the amplitude to peak at $\theta$, corresponding to a scattering amplitude of $\chi=\pi\mp\theta$. In the spinless (and non-magnetic) case with only orbital angular momentum, this also implies $\theta=\Delta\varphi$. 
\vspace{5pt}\\
Before moving on to the bulk of the paper let us briefly comment on notation, which can be at times tricky when taking the classical limit of quantum results. Throughout the paper, we keep factors of $\hbar$ explicit (while the speed of light is still $c=1$). Consequently, the different variables in our calculations have units that are a combination of [distance] and [mass]. The different quantities and their units are detailed in Table~\ref{tab:units}.

\begin{table}[ht]
\centering
\begin{tabular}{|c|c|c|}
\hline
{~~\textbf{Symbol}~~}          & {\textbf{Description}}                 & {\textbf{Units}} \\ \hline
$t$        & time                       & {[}distance{]}               \\ \hline
$r$        & distance                   & {[}distance{]}               \\ \hline
$m$        & mass                       & {[}mass{]}                   \\ \hline
$E$        & energy                     & {[}mass{]}                   \\ \hline
$k$        & momentum                   & {[}mass{]}                   \\ \hline
$\eta$        & Coulomb parameter                  & {[}mass{]}                   \\ \hline
$M$        & Schwarzschild mass                       & {[}mass{]}                   \\ \hline
$\ell$        & NUT parameter                       & {[}mass{]}                   \\ \hline
$j$        & angular momentum           & ~~{[}distance $\times$ mass{]}~~ \\ \hline
$\delta_j$ & phase shift                & ~~{[}distance $\times$ mass{]}~~ \\ \hline
$\nu$      & ~~effective angular momentum~~ & ~~{[}distance $\times$ mass{]}~~ \\ \hline
$h$      & ~~helicity~~ & ~~{[}distance $\times$ mass{]}~~ \\ \hline
$q$      & ~~pairwise helicity~~ & ~~{[}distance $\times$ mass{]}~~ \\ \hline
$\hbar$    & Planck's constant          & ~~{[}distance $\times$ mass{]}~~ \\ \hline
$G$    & Newton's constant          & ~~{[}distance $\times \,\text{mass}^{-1}${]}~~ \\ \hline
$~~\bar{j},\,\bar{\delta}_j,\,\bar{\nu},\,\bar{h},\,\bar{q}~~$ & $~~\hbar^{-1}j-\frac{1}{2},\,\hbar^{-1}\delta_j,\,\hbar^{-1}{\nu},\,\hbar^{-1}h,\,\hbar^{-1}q~~$ & dimensionless                \\\hline
$~~\bar{m},\,\bar{E},\,\bar{k},\,\bar{\eta}~~$ & $~~\hbar^{-1}m,\,\hbar^{-1}E,\,\hbar^{-1}k,\,\hbar^{-1}\eta~~$ & [$\text{distance}^{-1}$]                \\\hline
\end{tabular}
\caption{Different quantities used in their paper and their units when $\hbar\neq1,\,c=1$.}
\label{tab:units}
\end{table}

In our quantum calculation, we frequently make use of the dimensionless angular momentum variables
\begin{eqnarray}
&&~~~\bar{j}\,\equiv\,\hbar^{-1}j-\frac{1}{2},~~~\bar{\delta}_j\,\equiv\,\hbar^{-1}\delta_j,\nonumber\\[3pt]
&&\bar{\nu}\,\equiv\,\hbar^{-1}\nu,~~\bar{h}\,\equiv\,\hbar^{-1}h~~,~~\bar{q}\,\equiv\,\hbar^{-1}q\,.
\end{eqnarray}
In particular, $\bar{j},\,\bar{h}$ and $\bar{q}$ are quantized in half integer units. It is worth noting the $-\frac{1}{2}$ subtraction in the definition of $\bar{j}$, which corresponds to the famous Langer correction to the quantum angular momentum \cite{Berry_1972,Langer}. We also make use of the quantities
\begin{eqnarray}
&&\bar{m}\,\equiv\,\hbar^{-1}m,~~\bar{k}\,\equiv\,\hbar^{-1}k,\nonumber\\[3pt]
&&\bar{E}\,\equiv\,\hbar^{-1}E,~~\bar{\eta}\,\equiv\,\hbar^{-1}\eta\,,
\end{eqnarray}
all with units of [$\text{distance}^{-1}$].

\section{Semiclassical Scattering - Scalar Case}\label{sec:scalar}
In this section we derive the relation \eqref{eq:relation} by Poisson summation of the partial-wave expanded scattering amplitude in the classical limit \cite{Berry_1972}. Here start with the spinless, non-magnetic case (see also \cite{Damour:2019lcq, Bern:2020gjj,Gaddam:2020mwe,Gaddam:2020rxb} for related work), but will be able to generalize it to all spins and even to electric-magnetic scattering in the following sections. 

The partial-wave decomposed amplitude for $2\rightarrow 2$ scalar scattering takes the form
\begin{eqnarray}\label{eq:scalarJW}
\mathcal{A}=\frac{\hbar}{k} ~\sum_{\bar{j}=0}^{\infty}(2 \bar{j}+1) \frac{e^{2 i \bar{\delta}_{\bar{j}}}-1}{2 i} P_{\bar{j}}(\cos \theta),\,
\end{eqnarray}
where the $P_{\bar{j}}(x)$ are Legendre polynomials. Note that the map to effective one body dynamics here is trivial if we interpret $\mathcal{A}$ as the amplitude for an incoming plane wave to scatter to an angle $\theta$. Here and below, an overbarred quantities are quantum, for example, $\bar{j}=\hbar^{-1}j+\frac{1}{2}$ is a half integer quantum number.

We can write this expansion more compactly as
\begin{eqnarray}
\mathcal{A}=\sum_{\bar{j}}\,G(\bar{j})\,,
\end{eqnarray}
where 
\begin{eqnarray}
G(\bar{j})\equiv\frac{\hbar}{2ik}(2 \bar{j}+1)~e^{2 i \bar{\delta}_{\bar{j}}} ~P_{\bar{j}}(\cos \theta)\,.
\end{eqnarray}
We also drop the $-1$ part responsible for forward scattering, which does not survive in the classical limit. Using Poisson summation, we can express the same sum as
\begin{eqnarray}
\mathcal{A}=\sum_{n=-\infty}^{\infty}\,\int_0^\infty\,d\bar{j}\,e^{-2\pi i n j}\,G(j)\,.
\end{eqnarray}
Changing variables as $\bar{j}=\frac{j}{\hbar}-\frac{1}{2}$, we get
\begin{eqnarray}\label{eq:poisson}
\mathcal{A}=-\frac{i}{\hbar k}\,\sum_{n=-\infty}^{\infty}\,e^{-i\pi n}
\int_0^\infty\,dj\,j\,e^{\frac{2 i}{\hbar}\left[ \delta_{j}-\pi n j\right]}~P_{\bar{j}}(\cos \theta)\,.
\end{eqnarray}
Taking the leading term in $\hbar$ for $P_{\bar{j}}(\cos \theta)$ with $\bar{j}=\frac{j}{\hbar}-\frac{1}{2}$, we have for $\sin\theta \gtrsim \bar{j}^{\,-1}$\cite{Ford1959},
\begin{eqnarray}\label{eq:Papprox}
P_{\bar{j}}\left(\cos\theta\right)~\approx~\begin{cases}\left[\frac{j \pi}{2\hbar} \sin \theta\right]^{-1 / 2} \sin \left[\frac{j\theta}{\hbar}+\pi / 4\right]
\end{cases}\,.
\end{eqnarray}
Plugging this in \eqref{eq:poisson}, we have
\begin{eqnarray}\label{eq:poisson2}
\mathcal{A}=-\frac{1}{ k}\,\sqrt{\frac{1}{2\hbar\pi\sin \theta}}\sum_{n=-\infty}^{\infty}\,e^{-i\pi n}\left(e^{\frac{i\pi}{4}}I_+-e^{-\frac{i\pi}{4}}I_-\right)\,,
\end{eqnarray}
where
\begin{eqnarray}
I_\pm=\int_{0}^\infty\,dj\,\sqrt{j}\,e^{\frac{i}{\hbar}\left[ 2\delta_{j}+(\pm\theta-2\pi n )j\right]}\,.
\end{eqnarray}
In the $\hbar\rightarrow 0$ limit, we can take the saddle point approximation and get
\begin{eqnarray}\label{eq:poirel0}
\Theta\,\equiv\,\pm\theta-2\pi n\,=\,2\frac{d\delta_{j}}{dj}\,,
\end{eqnarray}
where $\Theta$ is the \textit{deflection} angle \cite{Berry_1972} and $n$ and the sign are fixed so that $0\leq\theta<\pi$ and $\Theta$ is positive for net repulsion. This in particular means that only one saddle point is selected for every value of $j$. The relation between the deflection angle and the \textit{scattering angle} $\chi$ is then
\begin{eqnarray}\label{eq:poirelch}
\chi\,=\,\pi\,-\,\Theta\,=\,\pi- 2\frac{d\delta_{j}}{dj}\,,
\end{eqnarray}
where the $\pm$ is 
Note that in the absence of spin or electric-magnetic angular momentum, the angular momentum $\vec{J}$ is just the orbital part. In this case, the motion is always in the plane transverse to $\vec{J}$, and we have $\chi=\Delta\varphi$. See \cite{Damour:2019lcq,Bern:2020gjj} for similar derivations of the relation \eqref{eq:poirelch}, in the absence of winiding. 

The interpretation of \eqref{eq:poirelch} is clear - the classical trajectory winds $n$ times and goes to infinity, after accumulating a total scattering angle $\chi=\Delta\phi$. The meaning of $\delta_{j}$ in the classical limit becomes even clearer once we consider the classical expression for the (azimuthal) scattering angle in terms of the radial action:
\begin{eqnarray}\label{eq:radac}
\Delta\varphi=-2\,\lim_{r\rightarrow\infty}\frac{d I_r(r)}{dj}\,.
\end{eqnarray}
See, for example \cite{landau1982mechanics} for a textbook derivation of this relation from the Hamilton-Jacobi equation, or \cite{Carter:1968ks} for the Hamilton-Jacobi equation for general geodesic motion. In light of this relation, we can make the satisfying identification \eqref{eq:relation}. This relation between the classical limit of the quantum phase shift and the integral of the radial action is not a coincidence. In fact, it is a direct consequence of the WKB approximation for $\delta_{j}$, which becomes \textit{exact} in the $\hbar\rightarrow0$ limit. We demonstrate this point in the next section by explicitly working out the phase shift for relativistic Coulomb scattering.

Taking the stationary phase approximation in \eqref{eq:poisson2}, we get the saddle point amplitude
\begin{eqnarray}\label{eq:ampradaccor}
i\mathcal{A}\,\sim\,e^{\frac{i}{\hbar}\left[\lim\limits_{r\rightarrow\infty}  \left(I_{r}(r)\,-\,d(r)\right)+\frac{\pi j}{2}-c(\hbar)\right]}\,,
\end{eqnarray}
which is valid in the $\hbar\rightarrow0$ limit. This is our saddle point version of the amplitude-angle relation \eqref{eq:ampradac}. 

\section{Relativistic Coulomb Scattering}\label{sec:Coulo}

To demonstrate the relations between the phase shift, the radial action and the classical scattering angle, we analyze the simple problem of a charged relativistic scalar scattering off a central Coulomb potential. The phase shifts in this problem are obtained by solving the Klein-Gordon (KG) equation for the scalar with mass $m$,
\begin{eqnarray}\label{eq:KGv}
\left[(\partial_\mu\,+\,i\hbar^{-1}\,Z\,e\,A_\mu)^2\,-\,\hbar^{-2}m^2\right]\Phi\,=\,\,0\,
\end{eqnarray}
in the background of the vector potential of a Coulomb charge: 
\begin{eqnarray}\label{eq:vecpocoul}
A_t=\frac{e}{4\pi r},\,\vec{A}\,=\,0\,.
\end{eqnarray}
Keeping in mind that we wish to eventually take the classical limit, we have made factors of $\hbar$ explicit in this equation. Taking $Z=-1$ for an attractive interaction and plugging in the vector potential, the KG equation takes the more familiar form
\begin{eqnarray}\label{eq:KGvst}
\left[\left(\partial_t\,-\,\frac{i\alpha}{\hbar r}\right)^2\,-\nabla^2\,\,-\,\hbar^{-2}m^2\right]\Phi\,=\,\,0\,.
\end{eqnarray}
As is standard in QM scattering problems, we solve the KG equation subject to regularity at $r=0$.
Substituting the ansatz \begin{eqnarray}
\Phi=e^{-i\bar{E}t}\,r^{-1}\,\sum_{\bar{j}}\,(2j+1)\,R_{\bar{j}}(r)\,P_{\bar{j}}(\cos\theta)
\end{eqnarray}
in \eqref{eq:KGv}, we get the radial equation
\begin{eqnarray}\label{eq:radcoul}
\hbar^2r^2\partial^2_r\,R_{\bar{j}}\,+\,\left[k^2\, r^2\,+\,2\eta k r\,-\,\hbar^2\bar{j}(\bar{j}+1)+\alpha\right]\,R_{\bar{j}}\,=\,0\, ,
\end{eqnarray}
where we defined
\begin{eqnarray}
k\,&=&\,\sqrt{E^2-m^2}\nonumber\\ \eta\,&=&\,E\alpha/k\,.
\end{eqnarray}
We can go further by defining  $\nu(\nu+\hbar)=\hbar^2\bar{j}(\bar{j}+1)-\alpha^2$ and so $\nu=\sqrt{j^2-\alpha^2}-\frac{\hbar}{2}$. In this case, the radial equation reduces to the form of a Coulomb equation,
\begin{eqnarray}\label{eq:radcoul2}
r^2\partial^2_r\,R_{\bar{j}}\,+\,\left[\bar{k}^2\, r^2\,+\,2\bar{\eta}\bar{k} r\,-\,\bar{\nu}(\bar{\nu}+1)\right]\,R_{\bar{j}}\,=\,0\, ,
\end{eqnarray}
where for any quantity $a$, the corresponding barred quantity is $\bar{a}\equiv\hbar^{-1}a$. The solution which converges at $r\rightarrow0$ is the Coulomb wavefunction (see \cite{NIST:DLMF}, equation 3.2.33):
\begin{eqnarray}\label{eq:coylwf}
R_{\bar{j}}\,=\,\frac{\Gamma(\bar{\nu}+1+i\bar{\eta})}{\Gamma(2\bar{\nu}+2)}\,M\left(-i\bar{\eta},\bar{\nu}+\frac{1}{2}~;~2i\bar{k}r\right)\, ,
\end{eqnarray}
where $M\left(\kappa,\mu;x\right)$ is the regular Whittaker function. As $r\rightarrow \infty$, the radial solution becomes
\begin{eqnarray}\label{eq:coylwfrl}
R_{\bar{j}}|_{r\rightarrow\infty}~\sim~e^{-i \left(\bar{k}r+\bar{\eta}\log(2\bar{k}r)-\frac{\pi \bar{j}}{2}\right)}\,+\,e^{2i\bar{\delta}_{j}}\,e^{i \left(\bar{k}r+\bar{\eta}\log(2\bar{k}r)-\frac{\pi\bar{j}}{2}\right)}\, ,
\end{eqnarray}
where
\[
2\delta_{j}=\pi(j-\nu-i\eta-\frac{\hbar}{2})-i\log\left\{\frac{\Gamma[\hbar^{-1}(\nu-i\eta)+1]}{\Gamma[\hbar^{-1}(\nu+i\eta)+1]}\right\}
\]
is related to the famous Coulomb phase shift. Taking the classical limit $\hbar\rightarrow 0$, we have
\begin{eqnarray}\label{eq:coulph}
2\delta_{j}\,=\,\pi(j-\nu) -2\nu\arctan\left(\frac{\eta}{\nu}\right)-2\eta\left[\log\left(\sqrt{\nu^2+\eta^2}\right)-1\right]+\eta\log(-i\hbar)\,.
\end{eqnarray}
By \eqref{eq:poirelch}, the classical scattering angle is then given by
\begin{eqnarray}\label{eq:coulang}
\chi=\Delta\varphi\,=\,\pi-2\frac{d\delta_{j}}{dj}\,=\,\frac{j}{\sqrt{j^2-\alpha^2}}\,\left[\pi+2\tan^{-1}\left(\frac{\alpha}{ \beta\sqrt{j^2-\alpha^2}}\right)\right]\,,
\end{eqnarray}
where $\beta=\frac{k}{E}$ is the velocity. This gives exactly the classical scattering angle for a relativistic scalar in a Coulomb potential \cite{Darwin1913,Boyer2004}-- the relativistic (scalar) generalization of the Rutherford scattering angle.
\\\quad\\
Finally, we can relate the phase shift \eqref{eq:coulph} to the one obtained in the WKB approximation \cite{dingle1973asymptotic}, which becomes exact in the classical limit. Starting from \eqref{eq:radcoul}, we substitue the WKB ansatz $R^\pm_{\bar{j}}\equiv r^{-1}e^{\pm i\frac{S(r)}{\hbar}}$ and get
\begin{eqnarray}\label{eq:WKBCoul}
&&\left[\partial_r S(r)\right]^2\,\mp\,i\hbar\partial^2_r S(r)\,=\,\mathcal{F}_{\text{Coul}}(r)+\hbar^2g_{\text{Coul}}(r)\nonumber\\[10pt]
&&\mathcal{F}_{\text{Coul}}(r)\equiv\frac{k^2r^2+2k\eta r-\nu^2}{r^2}~~,~~g_{\text{Coul}}(r)\equiv\frac{1}{4r^2}\,,
\end{eqnarray}
Expanding $S(r)$ and \eqref{eq:WKBCoul} to first order in $\hbar$, we get
\begin{eqnarray}\label{eq:WKBHJCoul}
S(r)\,&=&\,S_0(r)+\frac{i\hbar}{4}\log\left[\mathcal{F}_{\text{Coul}}(r)\right]\,+\,\mathcal{O}(\hbar^2)\nonumber\\[5pt]
\partial_r S_0(r)\,&=&\,\sqrt{\mathcal{F}_{\text{Coul}}(r)}\, .
\end{eqnarray}
The equation for $S_0$ is nothing but the radial Hamilton-Jacobi equation, and so $S_0(r)$ is simply equal to the radial action $I_r(r)$ in the classical limit. This is a well known fact about the WKB approximation. Equation \eqref{eq:WKBHJCoul} can be trivially integrated, so that
\begin{eqnarray}\label{eq:WKBCoulS}
I_r(r)=S_0(r)=\int_{r_\text{turn}}^r\,\sqrt{\mathcal{F}_{\text{Coul}}(r)}\,dr ,
\end{eqnarray}
where $r_{\text{turn}}=k^{-1}\left(-\eta+\sqrt{\eta^2+\nu^2}\right)$ is the classical turning point. The result is
\begin{eqnarray}\label{eq:WKBCoulSres}
I_r(r)=S_0(r)=\delta_{j}-\frac{\pi j}{2}+d(r)+c(\hbar)
\end{eqnarray}
where $c(\hbar)=-\frac{1}{2}\frac{\eta}{2}\log(-i\hbar)$. This is a further demonstration of the relation \eqref{eq:relation}. Clearly, this relation always holds as a consequence of two facts (A) the WKB approximation becomes exact in the classical limit (B) the WKB phase $S(r)$ coincides with the radial action $I_r(r)$ in the classical limit. 

Putting everything together, we get the WKB wavefunction using the ''right of barrier" linear combination of $R^\pm_{\bar{j}}\equiv r^{-1}e^{\pm i\frac{S(r)}{\hbar}}$, \cite{Ghatak}:
\begin{eqnarray}
R_{\bar{j}}\,&=&\,-\frac{i}{r{f^{\frac{1}{4}}_{\text{Coul}}(r)}}\,\left[e^{\frac{i}{\hbar}\left[I_r(r)+\frac{\pi}{4}+\mathcal{O}(\hbar^2)\right]}-e^{-\frac{i}{\hbar}\left[I_r(r)+\frac{\pi}{4}+\mathcal{O}(\hbar^2)\right]}\right]\,\nonumber\\[5pt]
&=&\,\frac{2}{r{f^{\frac{1}{4}}_{\text{Coul}}(r)}}\,\sin\left(\frac{1}{\hbar}\left[I_r(r)+\frac{\pi}{4}+\mathcal{O}(\hbar^2)\right]\right)\,.\nonumber\\
\end{eqnarray}

\begin{figure}[ht]
\begin{center}
\includegraphics[width=0.75\linewidth]{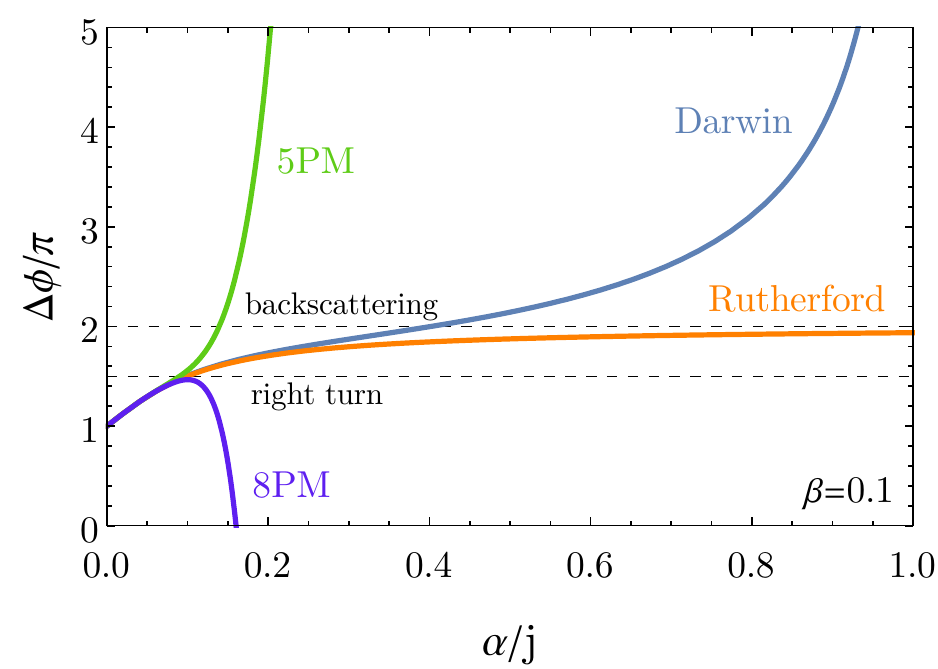}
\caption{Classical scattering angle $\Delta\varphi$ vs. $\alpha/j$ for relativistic Coulomb scattering. The result from eq. \eqref{eq:coulang} is shown in blue and labeled ``Darwin''. We also show the Ruthrford scattering angle in orange, as well as the expansions of \eqref{eq:coulang} to 5th and 8th orders in green and purple, respectively. We chose $\beta=0.1$ for this plot. Note that unlike Rutherford scattering, our classical trajectory involves winds around the origin for $\alpha/J\gtrsim0.4$. }\label{fig:Darwin}
\end{center}
\end{figure}

To get the scattering angle, we use the Hamilton-Jacobi relation 
\begin{eqnarray}\label{eq:WKBsolCoule}
\Delta\varphi=-2\lim_{r\rightarrow\infty}\frac{dI_r(r)}{dj}=\int_{r_{\text{turn}}}^\infty\,\frac{2j}{r^2\sqrt{\mathcal{F}_{\text{Coul}}(r)}}\,dr .
\end{eqnarray}
This integral can be carried out analytically, giving exactly the result \eqref{eq:coulang}. In figure~\ref{fig:Darwin} we plot this scattering angle (denoted by ``Darwin'', who was the first to derive it), together with the one for Rutherford scattering, $\Delta\varphi=\pi+2\arctan(\alpha/j)$. We also plot the expansions of \eqref{eq:coulang} to 5th and 8th order in $\alpha/j$. These clearly fail to converge beyond $\alpha/j\sim0.1$. At this point, the correct scattering angle is $\Delta\theta\sim\frac{3}{2}\pi$, so that the classical trajectory takes a right turn around the origin. This behaviour clearly is not well captured in perturbation theory, unless some sort of resummation is applied. 

In this vein, we wish to clarify a few important points about our formalism. Even though it involves a \textit{quantum} amplitude, it by no means relies on a \textit{perturbative expansion}. This is in contrast with other methods which rely on an expansion in $j^{-1}$. This is why we are able to capture classical effects that are inherently non-perturbative from the quantum point of view. The flip side, of course, is that we are working in the probe limit. We will come back to this point, and to future directions away from the probe limit, in the last section.

\section{Scalar in Schwarzschild Background}\label{sec:Schwarzschild}
As another application of our semiclassical analysis, we analyze the probe-limit scattering of a scalar in the background of a Schwarzschild BH. To find the phase shifts, we need to solve the Klein-Gordon equation subject to the boundary condition of an incoming wave \cite{Teukolsky:1973ha} at the BH horizon $r_h=2GM$, where $M$ is the Schwarzschild mass and $G$ is Newton's constant. The radial KG equation in a Schwarzschild background is \cite{Regge:1957td,Unruh:1976fm,Sanchez:1977si},
\begin{eqnarray}\label{eq:radSch}
\partial_r\left[\Delta\,\partial_r R\right]+\left[-\bar{j}(\bar{j}+1)+\frac{r^4E^2}{\hbar^2\Delta}-r^2\hbar^{-2}m^2\right]R\,=\,0\,.
\end{eqnarray}
with $\Delta=r\left(r-2GM\right)$. Again, we made factors of $\hbar$ explicit in this equation.

\subsection{WKB-Hamilton-Jacobi Analysis}
\begin{figure}[ht]
\begin{center}
\includegraphics[width=0.75\linewidth]{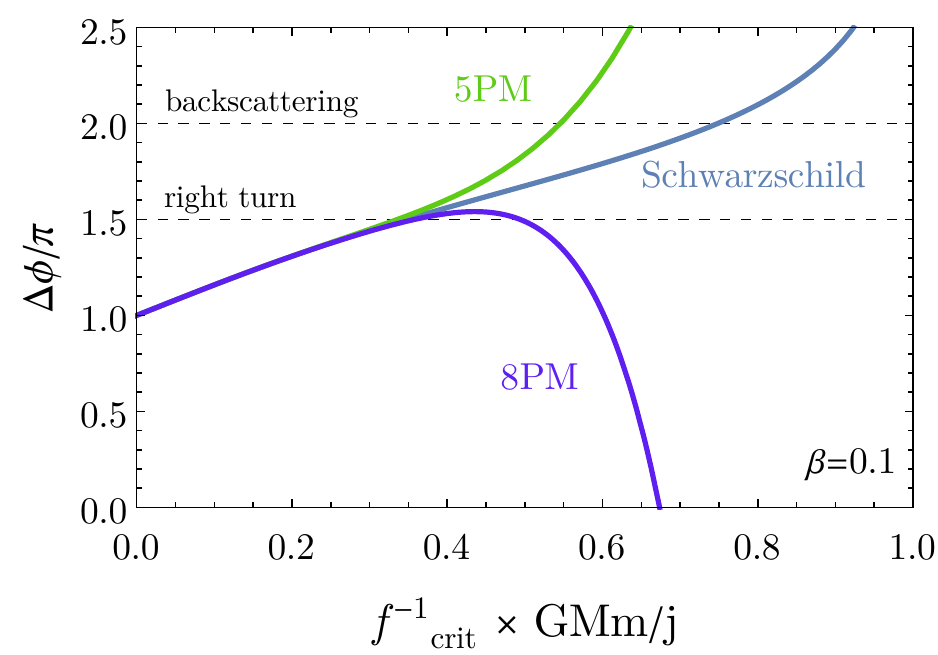}
\caption{Classical scattering angle $\Delta\varphi$ vs. $f^{-1}_{\text{crit}}GMm/j$, where $f_{\text{crit}}$ is the maximal allowed value of $GMm/j$ given by \eqref{eq:fcrit}. The result from eq. \eqref{eq:intsch} is shown in blue. Also shown are the 4PM and 8PM expansions of \eqref{eq:intsch} in green and purple, respectively. We chose $\beta=0.1$ for this plot. Our scattering angle corresponds to trajectories which wind around the origin for $GMm/j\gtrsim0.75f_{\text{crit}}$. }\label{fig:Sch}
\end{center}
\end{figure}

As a first step, we will extract the phase shifts from the WKB approximation to \eqref{eq:radSch}. This is guaranteed to reproduce the radial action and classical scattering angle, by virtue of the general map \eqref{eq:relation}. Substituting the WKB ansatz
\begin{eqnarray}
R^{\pm}_{\bar{j}}\,=\,\frac{1}{\sqrt{r(r-2GM)}}\,e^{\pm i\frac{S(r)}{\hbar}}\,,
\end{eqnarray}
we get
\begin{eqnarray}\label{eq:WKBHJSch}
&&~~~~~~~~~~~~~\left[\partial_r S(r)\right]^2\,\mp\,i\hbar\partial^2_r S(r)\,=\,\mathcal{F}_{\text{Sch}}(r)+\hbar^2g_{\text{Sch}}(r)\nonumber\\[10pt]
&&\mathcal{F}_{\text{Sch}}(r)\equiv\frac{k^2r^2+2k\xi r-\frac{r-2GM}{r}j^2}{(r-2GM)^2}~~,~~g_{\text{Sch}}(r)\equiv\frac{1}{4(r-2GM)^2}-\frac{GM}{2r^2(r-2GM)}\,,
\end{eqnarray}
where $\xi=\frac{GMm}{k}$. Expanding $S(r)$ and \eqref{eq:WKBHJSch} to first order in $\hbar$, we get
\begin{eqnarray}\label{eq:WKBHJSch0}
S(r)\,&=&\,S_0(r)+\frac{i\hbar}{4}\log\left[\mathcal{F}_{\text{Sch}}(r)\right]\,+\,\mathcal{O}(\hbar^2)\nonumber\\[5pt]
\partial_r S_0(r)\,&=&\,\sqrt{\mathcal{F}_{\text{Sch}}(r)}\, .
\end{eqnarray}
The equation for $S_0(r)$ is exactly the radial Hamilton-Jacobi equation \cite{Carter:1968ks,Carter:1968rr}, with $S_0(r)$ playing the role of the radial action
where $S_0(r)=I_r(r)$. The formal solution for this radial equation is then
\begin{eqnarray}\label{eq:WKBsolSch}
S_0\,=\,I_r(r)\,=\,\int_{r_\text{turn}}^r\,\sqrt{\mathcal{F}_{\text{Sch}}(r)}\,dr ,
\end{eqnarray}
where $r_{\text{turn}}$ is the largest real zero of $\mathcal{F}_{\text{Sch}}(r)$, corresponding to the classical turning point. Correspondingly, we get the ''right of barrier" \cite{Ghatak} WKB wavefunction
\begin{eqnarray}\label{eq:schwkb}
R_{\bar{j}}\,=\,\frac{1}{\sqrt{(r-2GM)}{f^{\frac{1}{4}}_{\text{Sch}}(r)}}\,\sin\left({\frac{1}{\hbar}\left[I_r(r)+\frac{\hbar}{4}+\mathcal{O}(\hbar^2)\right]}\right)\,.
\end{eqnarray}
To get the scattering angle, we again use the Hamilton-Jacobi relation 
\begin{eqnarray}\label{eq:WKBsolSche}
\Delta\varphi=-2\lim_{r\rightarrow\infty}\frac{dI_r(r)}{dj}=\int_{r_{\text{turn}}}^\infty\,\frac{2j}{r(r-2GM)\sqrt{\mathcal{F}_{\text{Sch}}(r)}}\,dr.
\end{eqnarray}
This integral can be carried out analytically, giving the all-order expression to the classical scattering angle of a probe scalar in a Schwarzschild background \footnote{Cf. \cite{Scharf} for an alternative expression in terms of elliptic functions.}
\begin{eqnarray}\label{eq:intsch}
     \Delta\varphi\,&=&\,\frac{4j}{k\,r_{\text{turn}}}~F_1\left(1;\frac{1}{2},\frac{1}{2};\frac{3}{2};\frac{r_{1}}{r_{\text{turn}}},\frac{r_{2}}{r_{\text{turn}}}\right)\,,
\end{eqnarray}
where $F_1$ is the Appell-$F_1$ function (see \cite{NIST:DLMF}, sec. 16.15).
The characteristic radii $r_{1,2}$ are defined by \begin{eqnarray}
p_{\text{Sch}}(r)\,\equiv\,\frac{r(r-2GM)^2}{k}\,\mathcal{F}_{\text{Sch}}(r)=(r-r_{\text{turn}})(r-r_{1})(r-r_{2})\,,
\end{eqnarray}
with $r_{1,2}\leq r_{\text{turn}}$, and the result is symmetric under $r_1\leftrightarrow r_2$. To have a sensible result, we need all three radii to be real, and so the cubic discriminant of $p_{\text{Sch}}(r)$ has to be non-negative. This, in turn, sets a lower bound on the angular momentum:
\begin{eqnarray}\label{eq:fcrit}
\frac{GMm}{J}\leq f_{\text{crit}}\equiv\frac{1}{4}\sqrt{\frac{1-18c_0-27c^2_0+\sqrt{(1+c_0)(1+9c_0)^3}}{2}}\,,~~~c_0=\frac{k^2}{m^2}\,.
\end{eqnarray}
Expanding \eqref{eq:intsch} to 4PM order, we get
\begin{eqnarray}\label{eq:intsch4PM}
    \frac{\Delta\varphi}{2}\,=\,\frac{\pi}{2}&+&\frac{2c_0+1}{\sqrt{c_0}}\left(\frac{GMm}{j}\right)+\frac{3 \pi \left(5 c_0+4\right)}{8}\left(\frac{GMm}{j}\right)^{2}\nonumber\\[5pt]
    &+&\frac{64c^3_0+72c^2_0+12c_0-1}{3{c}^{\frac{3}{2}}_0}\left(\frac{GMm}{j}\right)^{3}\nonumber\\[5pt]
    &+&\frac{105 \pi\left(33 c^2_0+48 c_0 +16\right)}{128}\left(\frac{GMm}{j}\right)^{4}+\mathcal{O}\left[{\left(\frac{GMm}{j}\right)}^5\right]\,,\nonumber\\
\end{eqnarray}
in complete agreement with \cite{Bjerrum-Bohr:2014zsa,Damour:2017zjx}. The full scattering angle, as well as its 5PM and 8PM expansions, are depicted in figure~\ref{fig:Sch}. As in the Coulomb case, the perturbative expansion fails at $GMm/j \sim 0.4f_{\text{crit}}$, which is approximately when the classical trajectory makes a full right turn around the BH.

\subsection{Full Quantum Solution and its Classical Limit}
\begin{figure}[ht]
\begin{center}
\includegraphics[width=0.75\linewidth]{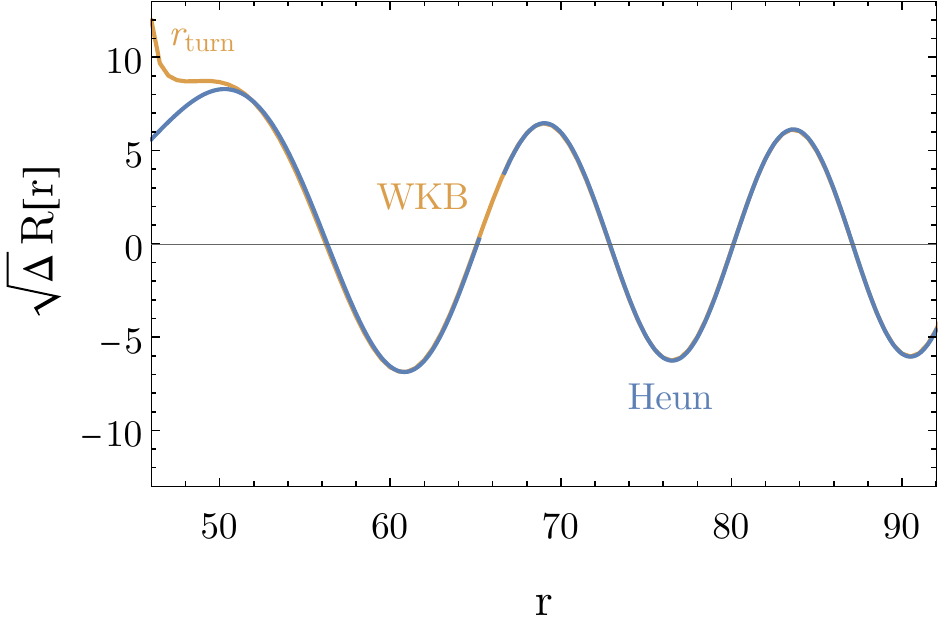}
\caption{Comparison of the full radial function and the WKB approximation for Schwarzschild scattering. The exact radial function is taken from eq.\eqref{eq:schfull}, while the WKB radial function is from eq. \eqref{eq:schwkb}. The WKB function diverges as usual at the classical turning point. Here $\Delta=r(r-2GM)$ and we've taken $M=1,\,m=0.1,k^2=0.1m^2,\,j=5$ and $\hbar=0.2$. Smaller values of $\hbar$ result in a numerical instability in the \textit{Mathematica} evaluation of HeunC. }\label{fig:schcomp}
\end{center}
\end{figure}

We now wish to relate the full quantum mechanical solution of \eqref{eq:radSch}, to the WKB-Hamilton-Jacobi result~\eqref{eq:WKBsolSch}.
The relevant boundary condition for this quantum  scattering problem is an incoming wave at the BH horizon~\cite{Teukolsky:1973ha}:
\begin{eqnarray}\label{eq:absbcsc}
R(r)|_{r\rightarrow2GM}~\sim~e^{-i\bar{E} r_*}~~,~~r_*=r+2GM\log\left(\frac{r-2GM}{2GM}\right)\,.
\end{eqnarray}
To this end we change variables to $z=1-\frac{r}{2GM}$ and substitute the ansatz $R=e^{-i\bar{E}r_*}e^{i(\bar{E}-\bar{k})(r-2GM)}H(z)$. Any solution with $H(2GM)=\text{const.}$ now satisfies the boundary condition \eqref{eq:absbcsc}. In terms of $H(z)$, the radial equation now becomes 
\begin{eqnarray}\label{eq:Heun}
z(z-1)H''(z)\,+\,\left[\bar{\gamma}(z-1)+\delta z+\epsilon z(z-1)\right]H'(z)+\left[\alpha z -q\right]H(z)=0\,,
\end{eqnarray}
where
\begin{eqnarray}
&&q=\bar{j}(\bar{j}+1)+2iGM(\bar{E}+\bar{k})-4{G^2M^2}(\bar{E}-\bar{k})^2,~~~~~\alpha=4iGM\bar{k}-4{G^2M^2}(\bar{E}-\bar{k})^2,\nonumber\\[5pt]
&&~~~~~~~~~~~~~~~~~~~~\bar{\gamma}=1-4iGM\bar{E},~~~~~~~~~~\delta=1,~~~~~~\epsilon=4iGM\bar{k}\,.
\end{eqnarray}
This equation is known as a \textit{Confluent Heun Equation}, and it has a solution
\begin{eqnarray}
H(z)=\HeunC\left(q,\alpha,\bar{\gamma},\delta,\epsilon;z\right)\, ,
\end{eqnarray}
which is defined in \textit{Mathematica12} and has a branch cut on the real line for $1\leq z <\infty$. The full solution to the radial equation is then
\begin{eqnarray}\label{eq:schfull}
R_{\bar{j}}(r)\,=\,e^{-i\bar{E}r_*}e^{i(\bar{E}-\bar{k})(r-2GM)}\,\HeunC\left(q,\alpha,\bar{\gamma},\delta,\epsilon;1-\frac{r}{2GM}\right)\, ,
\end{eqnarray}
and is regular for all $r>0$. Its asymptotic behavior is given by 
\begin{eqnarray}\label{eq:schwfrl}
R_{\bar{j}}|_{r\rightarrow\infty}~\sim~e^{-i \left(\bar{k}r+\bar{\eta}\log(2\bar{k}r)-\frac{\pi\bar{j}}{2}\right)}\,+\,e^{2i\bar{\delta}_{j}}\,e^{i \left(\bar{k}r+\bar{\eta}\log(2\bar{k}r)-\frac{\pi\bar{j}}{2}\right)}\, ,
\end{eqnarray}
where we defined the effective Coulomb parameter $\eta=GM\frac{E^2+k^2}{k}$, and also $\bar{\eta}=\hbar^{-1}\eta$.
\vspace{3pt}\\
The phase shifts $\bar{\delta}_j$ are not easy to obtain. For small $\bar{j}$, they can be obtained reliably using the well known method of Mano, Suzuki and Takasugi (MST) \cite{Mano:1996mf,Mano:1996vt}. In this method, several different solutions to \eqref{eq:Heun} are expressed as infinite sums of hypergeometric functions, following the seminal work of \cite{Leaver:1985}. A solution $H^0$ of \eqref{eq:Heun} that converges at $z=0$, is obtained as an infinite sum of Gauss Hypergeometric functions. $H^0$ is then expressed as a linear combination of the solutions $H^{\pm}$ that converge at $z\rightarrow\infty$, and are in turn given as an infinite sum of Coulomb functions. The MST method has been extensively used in the calculation of both BH perturbations and self-force corrections, see for example \cite{Berti:2006wq,Dolan:2008kf,Pan:2010hz,Bini:2013zaa,Bini:2014zxa,Damour:2016abl,Barack:2018yvs}
, as well as the living review \cite{Sasaki:2003xr}. The method is also implemented numerically in the Black hole Perturbation Toolkit code \cite{BHPToolkit}. However, in the $\hbar\rightarrow0$ limit we found it somewhat difficult to apply the MST method directly, and we leave it for future work. Instead, we simply plot the full solution \eqref{eq:schfull}, together with the WKB solution \eqref{eq:schwkb}. As can be seen from figure~\ref{fig:schcomp}, the two functions coincide already when we take $\hbar\sim0.2$. The spike of the WKB seen in the plot is the usual divergence at the classical at the classical turning point - which is usually resolved using an Airy function (see for example \cite{Ghatak}). Since we only care about the phase at $r\rightarrow\infty$, we will not dwell on this further.

\section{Generalization to Higher Spins and Electric-Magnetic Scattering}\label{sec:spinmag}
The expression \eqref{eq:scalarJW} can be easily generalized to the case of arbitrary spin particles by means of the Jacob-Wick formula
\footnote{See, for example~\cite{Haber:1994pe,Baratella:2020dvw} and chapter 2 of \cite{Pilkuhn:1979ps} for pedagogical presentations of the Jacob-Wick formalism. 
For an on-shell derivation of this formula, see \cite{Arkani-Hamed:2017jhn,Jiang:2020rwz}.} \cite{Jacob:1959at}. The $2\rightarrow2$ scattering amplitude for particles with helicities
\footnote{As the particles can be massive, their helicities are not Lorentz invariant. Here we specialize to the COM frame with the incoming particles traveling along the z-axis.} $\bar{h}_{1,..,4}$ is
\begin{eqnarray}\label{eq:JWn}
\mathcal{A}\,=\,N\,e^{i(\bar{h}_{12}-\bar{h}_{34})\varphi}\,\sum_{\bar{j}}\,(2\bar{j}+1) \frac{e^{2 i \bar{\delta}_{\bar{j}}}-1}{2 i}\,d^{\bar{j}}_{\bar{h}_{12},\bar{h}_{34}}(\theta)\,,
\end{eqnarray}
where $h_{1,..,4}$ are the helicities of the scattered particles, $h_{ab}\equiv h_a-h_b$. The normalization $N$ is conventionally taken as $N=\sqrt{8\pi s}$. Meanwhile, $d$ is the famous Wigner matrix, defined as 
\begin{eqnarray}
d^{\bar{j}}_{ab}(\theta)\,\equiv\,\bk{{\bar{j}},a|e^{-i\theta J_z}|{\bar{j}},b}\,.
\end{eqnarray}
For simplicity, we will focus on the case where $\bar{h}_{12}=a \bar{h}_{34}\equiv \bar{h}$, with $a=\pm$. In this case we have, up to a phase,
\begin{eqnarray}\label{eq:JWn2}
\mathcal{A}\,=\,N\,\sum_{\bar{j}}\,(2 \bar{j}+1) \frac{e^{2 i \bar{\delta}_{\bar{h}}}-1}{2 i}\,f_a\,d^{\bar{j}}_{\bar{h},a \bar{h}}(\theta)\,,~~~a=\pm\,
\end{eqnarray}
and $f_+=(-1)^{\bar{j}+\bar{h}},\,f_-=1$. The case $a = +$ in the equation above is easily interpreted in terms of effective one body dynamics: it is the amplitude for a helicity $h$ plane wave to scatter into the angle $\theta$. We will comment on the one body interpretation of $a=-$ below. 

Our next step is to take the classical limit, noting that $h=\hbar \bar{h}$ is the classical spin/helicity of the particle which is finite when $\hbar\rightarrow 0$. Repeating the Poisson summation of \eqref{eq:poisson}, we can take the classical limit of the $d$-matrix as worked out by Schwinger et al. in \cite{Schwinger:1976fr}. Defining $\theta_+=\pi-\theta$ and $\theta_-=\theta$, we have
\begin{eqnarray}\label{eq:dapprox}
f_a\,d^{\bar{j}}_{\bar{h},a \bar{h}}\left(\theta\right)~\rightarrow~\frac{\hbar^{\frac{1}{2}}}{\left[\cos^{2} \frac{\theta_a}{2}-\frac{h^2}{j^2}\right]^{1 / 4}\sqrt{\pi j \sin \frac{\theta_a}{2}}}\sin\left(\alpha_a \,\frac{j}{\hbar}-\beta_a \,\frac{h}{\hbar} +\frac{\pi}{4}\right),~~~~~\sin\theta \gtrsim \bar{j}^{\,-1}
\end{eqnarray}
Where $a=\pm$, $\sin\left(\frac{\alpha_{a}}{2}\right)={\left(1-\frac{h^2}{j^2}\right)}^{-\frac{1}{2}}\sin\left(\frac{\theta_a}{2}\right)$ and $\sin\left(\frac{\beta_a}{2}\right)={\frac{h}{j}\left(1-\frac{h^2}{j^2}\right)}^{-\frac{1}{2}}\tan\left(\frac{\theta_a}{2}\right)\,.$ 
For $h=0$ the Schwinger approximation reproduces \eqref{eq:Papprox}.
Plugging the asymptotic form \eqref{eq:dapprox} into the Poisson sum, we now get the generalization of \eqref{eq:poisson2},
\begin{eqnarray}\label{eq:Poiss2}
\mathcal{A}_{\bar{h},a\bar{h}}=-\frac{N}{\hbar ^{3/2}}\frac{1}{\sqrt{\pi\sin \frac{\theta_{a}}{2}}}\,\sum_{n=-\infty}^{\infty}\,e^{-i\pi n}
\left(e^{\frac{i\pi}{4}}I_+-e^{-\frac{i\pi}{4}}I_-\right)\,.
\end{eqnarray}
with
\begin{eqnarray}\label{eq:Poiss2int}
I_\pm\,=\,\int_{0}^\infty\,dj\,\frac{\sqrt{j}}{\left[\cos^{2} \frac{\theta_a}{2}-\frac{h^2}{j^2}\right]^{1 / 4}}e^{\frac{i}{\hbar}\left[2\delta_{j}\,\pm(\alpha_{a} \,j-\beta_{a} \,h)+2\pi n j\right]}\,.
\end{eqnarray}
Following Schwinger et al. \cite{Schwinger:1976fr}, we take the stationary phase approximation of $I_\pm$, and find that the classical scattering angle $\chi$ is given by
\begin{eqnarray}\label{eq:thetasol}
+\,\text{case:}~~~~\sin\left(\frac{\chi}{2}\right)\,&=&\,\sqrt{1-\frac{h^2}{j^2}}\,\cos\left(\frac{d\delta_{j}}{dj}\right)\nonumber\\
-\,\text{case:}~~~~\cos\left(\frac{\chi}{2}\right)\,&=&\,\sqrt{1-\frac{h^2}{j^2}}\,\sin\left(\frac{d\delta_{j}}{dj}\right)\,.
\end{eqnarray}
Note that in this case the classical trajectory is not confined to the $xy$-plane, and so $\chi\neq\Delta\varphi$. However, the relation $\Delta\varphi=-2\lim_{r\rightarrow\infty}\frac{dI_r(r)}{dj}=\pi-2\frac{d\delta_{j}}{dj}$ still holds as a consequence of the WKB-Hamilton-Jacobi analysis.

\subsection{Classical Interpretation}
To interpret \eqref{eq:thetasol}, we pick $a=+$ such that the effective one body dynamics is that of a spin $h\,\hat{p}$ particle scattering in a central field. The particle enters with its momentum along $\hat{p}_{\text{in}}$ and leaves along $\hat{p}_{\text{out}}$, with the scattering angle given by $\cos(\chi)=\hat{p}_{\text{in}}\cdot \hat{p}_{\text{out}}$. As the particle moves from $t\rightarrow-\infty$ to $t\rightarrow\infty$, it gets deflected azimuthally by $\Delta\varphi=\pi-2\frac{d\delta_{j}}{dj}$ in the plane transverse to $\vec{J}$, or in other words: 
\begin{eqnarray}\label{eq:Jtotrel2}
\cos\left(\Delta\varphi\right)\,&=&\,\frac{\vec{p}_{\text{in}\perp}\cdot \vec{p}_{\text{out}\perp}}{|\vec{p}_{\text{in}\perp}||\vec{p}_{\text{out}\perp}|}\nonumber\\[5pt]
\vec{p}_{i\perp}\,&\equiv&\,\hat{p}_{i}-\left(\hat{p}_{i}\cdot\hat{J}\right)\hat{J}\,.
\end{eqnarray}
In addition, the motion is constrained by total angular momentum conservation. The total angular momentum in this case is
\begin{eqnarray}\label{eq:Jtot}
\vec{J}=\vec{r}\times\vec{p}+h\,\hat{p}\,.
\end{eqnarray}
Since $\hat{p}=\mp\hat{r}$ at $t\rightarrow\pm\infty$, we have
\begin{eqnarray}\label{eq:Jtotrel}
\hat{p}_{\text{in}}\cdot\vec{J}=\hat{p}_{\text{out}}\cdot\vec{J}=h\,.
\end{eqnarray}
The combination of \eqref{eq:Jtotrel2} and \eqref{eq:Jtotrel}, and some elementary trigonometry, leads to
\begin{eqnarray}\label{eq:thetasol2}
\sin\left(\frac{\chi}{2}\right)\,&=&\,\sqrt{1-\frac{h^2}{j^2}}\,\sin\left(\frac{\Delta\varphi}{2}\right)\,,
\end{eqnarray}
consistently with \eqref{eq:thetasol}.

\subsection{Electric-Magnetic Scattering}
In \cite{Csaki:2020inw}, the Jacob-Wick formula was generalized to the $2\rightarrow2$ scattering of electric-magnetic scattering, i.e. the scattering of \textit{mutually non-local} particle like an electric charge and a monopole or two dyons with charges $(e_i,g_i)$. First, we define the \textit{pairwise helicity} \cite{Csaki:2020uun} of the two particles to be the half integer
\begin{eqnarray}\label{eq:pairwise}
\bar{q}=\frac{e_1 g_2-e_2 g_1}{\hbar}\,.
\end{eqnarray}
We also define the corresponding classical quantity $q=\hbar\bar{q}$. The generalized Jacob-Wick formula is then
\begin{eqnarray}\label{eq:JWmag}
\mathcal{A}^q\,&=&\,N\,\sum_{\bar{j}}\,(2 \bar{j}+1) \frac{e^{2 i \bar{\delta}_{\bar{j}}}-1}{2 i}\,\mathcal{D}^{\bar{j}*}_{\bar{q}+\bar{h}_{12},-\bar{q}+\bar{h}_{34}}(\varphi,\theta,-\varphi)\nonumber\\
&=&\,N\,e^{i(2\bar{q}+\bar{h}_{12}-\bar{h}_{34})\varphi}\,\sum_{\bar{j}}\,(2 \bar{j}+1) \frac{e^{2 i \bar{\delta}_{\bar{j}}}-1}{2 i}\,d^{\bar{j}}_{\bar{q}+\bar{h}_{12},-\bar{q}+\bar{h}_{34}}(\theta)\,.
\end{eqnarray}
The $\bar{q}-$modified $\mathcal{D}$-matrices are also known as \textit{monopole harmonics} \cite{Wu:1976ge} or \textit{spin-weighted spherical harmonics} \cite{Teukolsky:1973ha}.
Their appearance reflects the fact that the total angular momentum $\vec{J}$ includes a contribution $-q\hat{r}$ from the electromagnetic field sourced by the scattering dyons. The generalized Jacob-Wick formula \eqref{eq:JWmag} was derived in \cite{Csaki:2020inw} for fermion-monopole scattering, but it can be shown to hold for all helicities by the same pairwise helicity arguments.

The semiclassical result \eqref{eq:thetasol} holds in this case as well. In fact, to obtain the scattering angle for a scalar charge on a scalar monopole, we simply take $h_{12}=h_{34}=0$ but $q=\hbar \bar{q}\neq0$. We can now directly apply \eqref{eq:thetasol} with $a=-$ and $h=q$ and get \cite{Boulware:1976tv,Schwinger:1976fr} 
\begin{eqnarray}\label{eq:thetasolmag}
\cos\left(\frac{\chi}{2}\right)\,=\,\sqrt{1-\frac{q^2}{j^2}}\,\cos\left(\frac{\Delta\varphi}{2}\right)\,.
\end{eqnarray}
The classical interpretation of this relation is similar to the $a=+$ case, with a slight modification. The total angular momentum is now given by
\begin{eqnarray}\label{eq:Jtotmon}
\vec{J}=\vec{r}\times\vec{p}-q\,\hat{r}\,,
\end{eqnarray}
which means that the entire motion is confined to the cone $\hat{r}\cdot\vec{J}=-q$. Together with \eqref{eq:Jtotrel2}, this immediately gives the result \eqref{eq:thetasolmag}.

We can now apply the semiclassical limit to charge-monopole scattering, or equivalently to its gravitational double copy, a probe mass in Newman-Taburino-Unti (NUT) space. We will do this in the next two sections.

\section{Charge-Monopole Scattering}\label{sec:monopole}

By the arguments of the previous section, the classical scattering angle for a scalar charge in the background of a scalar monopole is given by
\eqref{eq:thetasolmag}, with $\left|\Delta\varphi\right|=-2\frac{d\delta}{dj}$. To apply it, we need to solve for the phase shifts of a scalar plane wave in the background field of the monopole. To do this, we solve the Klein-Gordon equation \eqref{eq:KGv}, with a vector potential given by 
\begin{eqnarray}\label{eq:vecpomon}
A_t=0~~,~~\vec{A}=g\frac{1-\cos\theta}{r\sin\theta}\,\hat{\varphi}\,,
\end{eqnarray}
and $\hat{\varphi}=-\sin\varphi\hat{x}+\cos\varphi\hat{y}$. This potential has a ``Dirac string'' along the negative $\hat{z}$ axis,
but we will see without loss of generality that the test charge always stays in the upper hemisphere, so this will not play a role in our analysis. Of course the formal way to fix this is to define the vector potential on the north and south hemispheres separately \cite{Wu:1976ge}.

Substituting the vector potential in the KG equation \eqref{eq:KGv}, we find 
\begin{eqnarray}\label{eq:KGL}
\hbar^2\partial^2_t\Phi\,-\left[\hbar^2r^{-2}\partial_r(r^2\,\partial_r)\,- \,\frac{\hbar^{-2}J^2\,-\,q^2}{r^2}\,-\,m^2\,\right]\,\Phi\,=\,0\,.
\end{eqnarray}
Here the squared angular momentum operator $J^2$ is given by \cite{Shnir:2005xx}
\begin{eqnarray}\label{eq:angmom}
J^2\,=\,\bar{q}^2-\frac{1}{\sin ^{2} \theta}\left[\sin \theta \frac{\partial}{\partial \theta}\left(\sin \theta \frac{\partial}{\partial \theta}\right)\,+\,\left(\frac{\partial}{\partial \varphi}\,-\,i \bar{q}\,(1-\cos \theta)\right)^{2}\right]\,.
\end{eqnarray}
As expected, this operator is modified by the presence of the angular momentum $-q\hat{r}$ carried by the EM field. The eigenvalues of $J^2$ are related to our $d$-matrices
\vspace{4pt}
\begin{eqnarray}\label{eq:eigv}
J^2\,\mathcal{D}^{\bar{j}*}_{\bar{q},-\bar{q}}(\varphi,\theta,-\varphi)\,=\,\bar{j}(\bar{j}+1)\,\mathcal{D}^{\bar{j}*}_{\bar{q},-\bar{q}}(\varphi,\theta,-\varphi)\,,
\end{eqnarray}
where $\mathcal{D}^{\bar{j}*}_{\bar{q},-\bar{q}}(\varphi,\theta,-\varphi)=e^{2iq\varphi}\,d^{\bar{j}}_{\bar{q},-\bar{q}}(\theta)$ are the complex conjugates of Wigner's $\mathcal{D}$-matrices.
\vspace{4pt}
Now we can separate variables as $\Phi=e^{-i\bar{E}t}\,R(r)\,e^{2i\varphi}\,d^{\bar{j}}_{\bar{q},-\bar{q}}(\theta)$ and get the radial equation
\begin{eqnarray}\label{eq:sphbes}
\left[\hbar^2r^2\partial^2_r\,+\,2\hbar^2r\partial_r\,+\,k^2\, r^2\,-\,\nu(\nu+\hbar)\right]\,R\,=\,0\, ,
\end{eqnarray}
where $k^2=E^2-m^2$ and
\begin{eqnarray}\label{eq:ltilde}
\nu\,&\equiv&\,\hbar\bar{\nu}\nonumber\\
\bar{\nu}(\bar{\nu}+1)\,&\equiv&\,\bar{j}(\bar{j}+1)-\bar{q}^2\rightarrow\nonumber\\[5pt]
\nu\,&\equiv&\,\sqrt{j^2\,-\,q^2}-\frac{\hbar}{2}\,.
\end{eqnarray}
We recall here the Langer correction \cite{Langer} of $-\frac{1}{2}$ in $j=\hbar^{-1}\bar{j}-\frac{1}{2}$. Equation \eqref{eq:sphbes} a spherical Bessel equation whose regular solution at $r\rightarrow 0$ is $R(r)\,=\,j_{\bar{\nu}}(\bar{k} r)$. Asymptotically we have
\begin{eqnarray}\label{eq:coylwfrlmon}
R|_{r\rightarrow\infty}~\sim~e^{-i(\bar{k}r-\frac{\pi\bar{j}}{2})}\,+\,e^{2i\bar{\delta}_{j}}\,e^{i(\bar{k}r-\frac{\pi\bar{j}}{2})}\, ,
\end{eqnarray}
with $2\delta_{j}=\pi(j-\nu-\frac{\hbar}{2})$. Taking the classical limit, we have
\begin{eqnarray}\label{eq:monph}
2\delta_{j}\,=\,\pi\,(j-\sqrt{j^2-q^2})\,
\end{eqnarray}
By \eqref{eq:thetasolmag}, the classical scattering angle is then given by
\begin{eqnarray}\label{eq:monang}
\cos\left(\frac{\chi}{2}\right)\,=\,\sqrt{1-\frac{q^2}{j^2}}\,\cos\left(\frac{\pi}{2\sqrt{1-\frac{q^2}{j^2}}}\right)\,.
\end{eqnarray}

This all order expression was first derived in \cite{Banderet:1946fm,Schwinger:1976fr,Boulware:1976tv} in a non-relativistic context, and it exactly reproduces the classical calculation of \cite{Schwinger:1976fr,Boulware:1976tv}.

\subsection{Probe mass in NUT Space}\label{sec:NUT}
In this section we use our phase shift formalism to derive the all-order classical scattering angle for a probe mass (equivalently, a probe Schwarzschild BH) in the background of a pure NUT, i.e. the $M_{\text{NUT}}\rightarrow0$ limit \cite{bonnor1969} of Taub-NUT space \cite{Taub:1950ez,NUT}. NUT space is the double copy of a magnetic monopole \cite{Luna:2015paa,Caron-Huot:2018ape,Huang:2019cja,Alawadhi:2019urr,Kol:2020ucd,Moynihan:2020gxj,Emond:2020lwi,Kim:2020cvf,Alawadhi:2021uie}, and so this problem is the double copy of charge-monopole scattering. 

The metric for Taub-NUT space in Boyer-Lindquist coordinates is given by 
\vspace{2pt}
\begin{eqnarray}\label{eq:taubmet}
ds^2_{\text{Taub-NUT}}\,=\,-f(r)\,\left[dt+2G\ell(\cos\theta-1)\,d\varphi\right]^2\,+\,\frac{1}{f(r)}\,dr^2\,+\,\left(r^2+G^2\ell^2\right)\left(d\theta^2\,+\,\sin^2\theta\,d\varphi^2\right)\, ,
\end{eqnarray}
where $M$ is the mass and $G\ell$ in the NUT charge while 
\[
f(r)= \frac{r^2-2GMr-G^2\ell^2}{r^2+G^2\ell^2} \,.
\]
Here we use the upper hemisphere metric of \cite{Kol:2020ucd}. The pure NUT metric is then obtained by setting $M=0$ (conversely, in the $\ell=0$ limit the metric reduces to the Schwarzschild metric). 

As we will show explicitly below, when considering the scattering of a test mass $\Phi$ in NUT space, regularity of the angular wavefuctions at the Misner string \cite{Misner:1963fr} (here on the negative $z$-axis) requires the quantization of
\begin{eqnarray}
\bar{q}\,\equiv\,\hbar^{-1}q\,\equiv\,\hbar^{-1}(2G\ell E)\,,
\end{eqnarray} 
in half-integer units (see also \cite{Dowker1974} for a similar conclusion). This is a further validation of the classical double copy relation between mass-NUT scattering and charge-monopole scattering \cite{Luna:2015paa,Huang:2019cja,Alawadhi:2019urr,Kol:2020ucd,Moynihan:2020gxj,Emond:2020lwi,Kim:2020cvf,Alawadhi:2021uie} - this time in terms of an inherently non-perturbative quantization condition. In complete analogy with the charge-monopole case, the gravitational field sourced by the NUT and the probe mass contains additional angular momentum $-q\hat{r}$, where $\hat{r}$ is the unit vector from the NUT to the probe mass. This has been shown classically a long time ago \cite{bonnor1969,Dowker1974,Zimmerman1989,Kagramanova:2010bk,Clement:2015cxa,Frolov:2017kze}.

Note that this angular momentum is proportional to the overall \textit{energy} of the probe mass and not its \textit{rest mass}, which we can take to be small or even zero. In other words, this is \textit{not} a gravitational backreaction effect from the mass of the probe, but rather the net angular momentum carried by the soft gravitons exchanged between the NUT and the probe. 

Consequently, the modified Jacob-Wick formula \eqref{eq:JWmag} is also valid in the mass-NUT case, with appropriate phase shifts $\delta_j$. To compute these phase shifts, we solve the Klein-Gordon equation 
\begin{eqnarray}
\left(D_\mu D^\mu\,-\,\hbar^{-2}m^2\right)\,\Phi\,=\,0\, ,
\end{eqnarray}
in the background \eqref{eq:taubmet}. Writing the D'Alembertian explicitly and substituting the ansatz $\Phi=e^{-i\bar{E} t}\,T(r,\theta,\phi)$, we get
\begin{eqnarray}\label{eq:eqTaub}
\hbar^2(r^2+G^2\ell^2)^{-1}\partial_r\left[\left(r^2-G^2\ell^2\right)\partial_r T\right]\,+\,\left[\frac{r^2+G^2\ell^2}{r^2-G^2\ell^2}E^2-\frac{\hbar^{2}J^2-q^2}{r^2+G^2\ell^2}-m^2\right]T\,=\,0\, .
\end{eqnarray}
Here $k^2=E^2-m^2, \,q=2G\ell E$, and the squared angular momentum operator $J^2$ is given by the same expression as the monopole case, \eqref{eq:angmom}, whose eigenfunctions are $\mathcal{D}^{\bar{j}*}_{\bar{q},-\bar{q}}(\varphi,\theta,-\varphi)=e^{2iq\varphi}\,d^{\bar{j}}_{\bar{q},-\bar{q}}(\theta)$. Now we can separate variables as
\begin{eqnarray}
T(r,\theta)\,=\,\sum_{\bar{j}}(2\bar{j}+1)\,R_{\bar{j}}(r)\,e^{2iq\varphi}\,d^{\bar{j}}_{\bar{q},-\bar{q}}(\theta)\,,
\end{eqnarray}
and find (cf. \cite{Dowker1974,Bini:2003sy})
\begin{eqnarray}\label{eq:eqTaubr}
\hbar^2(r^2-G^2\ell^2)^{-1}\partial_r\left[\left(r^2-G^2\ell^2\right)\partial_r R_{\bar{j}}\right]\,+\,\left[\frac{r^2+G^2\ell^2}{r^2-G^2\ell^2}E^2-\frac{j^2-\frac{\hbar^2}{2}-q^2}{(r^2-G^2\ell^2)}-m^2\right]R_{\bar{j}}\,=\,0\, .
\end{eqnarray}
We will now solve this equation and extract the classical scattering angle in two ways: first, by the WKB-Hamilton-Jacobi method, followed by taking the classical limit of the full quantum solution.

\subsection{WKB-Hamilton-Jacobi Analysis}
Substituting the WKB ansatz  $R_{\bar{j}}=(r^2-G^2\ell^2)^{-1/2}\,e^{i\hbar^{-1}I_r(r)}$, we deduce the radial Hamilton-Jacobi equation \cite{Carter:1968rr,Zimmerman1989,Kagramanova:2010bk,Clement:2015cxa,Frolov:2017kze}. To leading order in $\hbar$, this is 
\begin{eqnarray}\label{eq:WKBHJNUT}
\partial_r I_r(r)\,=\,\sqrt{{\mathcal{F}_{\text{NUT}}}(r)},~~~{\mathcal{F}_{\text{NUT}}}(r)\equiv\frac{r^2+G^2\ell^2}{r^2-G^2\ell^2}\left(\frac{r^2+G^2\ell^2}{r^2-G^2\ell^2}E^2-\frac{j^2-q^2}{r^2+G^2\ell^2}-m^2\right)\, ,
\end{eqnarray}
The formal solution for this radial equation is then
\begin{eqnarray}\label{eq:WKBsolNUT}
I_r(r)=\int_{r_\text{turn}}^r\,\sqrt{{\mathcal{F}_{\text{NUT}}}(r)}\,dr ,
\end{eqnarray}
where $r_{\text{turn}}$ is the largest real zero of ${\mathcal{F}_{\text{NUT}}}(r)$, corresponding to the classical turning point. To determine the scattering angle, we use the Hamilton-Jacobi relation
\begin{eqnarray}\label{eq:WKBsolNUTe}
\Delta\varphi=-2\lim_{r\rightarrow\infty}\frac{dI_r(r)}{dj}=\int_{r_{\text{turn}}}^\infty\,\frac{2j}{(r^2-G^2\ell^2)\sqrt{{\mathcal{F}_{\text{NUT}}}(r)}}\,dr\, .
\end{eqnarray}
The above integral can be carried out analytically, giving the all-order expression to the classical scattering angle of a probe scalar in a NUT background 
\begin{eqnarray}\label{eq:intNUT}
     \Delta\varphi\,&=&\,\frac{2j}{k r_+}~K\left(\frac{r_-}{r_+}\right)\,,
\end{eqnarray}
where $K(k_{\text{mod}})$ is Legendre's complete elliptic integral, and in this case it is a function of $k_{\text{mod}}\equiv r_-/r_+$. Note that in \textit{Mathematica}, elliptic integrals are expressed as functions of $m_{\text{mod}}=k^2_{\text{mod}}$. The characteristic radii $r_\pm$ are defined by
\[
\frac{(r-G\ell)^2}{k}\,{\mathcal{F}_{\text{NUT}}}(r)=(r^2-r^2_+)(r^2-r^2_-) \,,
\]
with $r_-\leq r_+\equiv r_{\text{turn}}$.
Explicitly, $r_\pm$ are 
\begin{eqnarray}
r_\pm\,=\,\frac{j}{\sqrt{2}k}\,\sqrt{1-\frac{3}{2}\frac{q^2}{j^2}\pm\sqrt{1-(\beta^2+3)\frac{q^2}{j^2}+\left[\frac{1}{4}(\beta^2+1)+2\right]\frac{q^4}{j^4}}}\,,~~~~~\beta=k/E\,.
\end{eqnarray}

Expanding $\Delta\varphi$ to 6PM order, we find
\begin{eqnarray}\label{eq:NUTphi6PM}
\Delta\varphi\,=\,\pi&+&\frac{3 \pi   \left(5
  E^2-m^2\right)}{16
  E^2}{\left(\frac{2G\ell E}{j}\right)}^2+\frac{3 \pi 
  \left(515 E^4-246
  E^2 m^2+19 m^4\right)}{1024 E^4
  }{\left(\frac{2G\ell E}{j}\right)}^4\nonumber\\[5pt]
  &+&\frac{15 \pi
  \left(3257 E^6-2599 E^4
  m^2+515 E^2 m^4-21 m
  ^6\right)}{16384 E^6
  }{\left(\frac{2G\ell E}{j}\right)}^6\,+\,\ldots
\end{eqnarray}
It is nice to note that this expression is consistent with the 2PM result obtained by on-shell methods in \cite{Kim:2020cvf}. This resolves the apparent discrepancy in \cite{Kim:2020cvf} in a simple way - by evaluating the integral \eqref{eq:WKBsolNUTe} analytically as an elliptic integral.

\subsection{Full Quantum Solution and its Classical Limit}
Here we find the exact phase shifts of the full quantum problem, and show that their classical limit reproduces \eqref{eq:intNUT}. Changing variables in \eqref{eq:eqTaubr} as $R_{\bar{j}}(r)=F(x)$ with $x=\frac{2Er}{q}$, and remembering that $q=2G\ell E$, we deduce the radial equation
\begin{eqnarray}\label{eq:radTaubsph}
\partial_x\left[\left(1-x^2\right)\partial_x F\right]+\left[\bar{\lambda}+\bar{\gamma}^2(1-x^2)-\frac{\bar{\mu}^2}{1-x^2}\right]F\,=\,0
\end{eqnarray}
where
\begin{eqnarray}\label{eq:lamb}
\hbar^2\bar{\lambda}\,=\,j^2-\frac{\hbar^2}{4}-\frac{q^2}{2}\left(\beta^2+3\right),~~\hbar\bar{\gamma}\,=\,\frac{\beta q}{2}, ~~\hbar\bar{\mu}\,=\,iq\,,
\end{eqnarray}
and $\beta=k/E$. This is a prolate spheroidal equation with a complex separation constant $\bar{\mu}$ - a very well studied equation \cite{meixner2006mathieu,arscott1964periodic,Falloon_2003}. The solution we are looking for is the one which satisfies an absorbing boundary condition at $r=G\ell$ \cite{Teukolsky:1973ha}:
\begin{eqnarray}\label{eq:absbc}
F(x)|_{x\rightarrow1}~\sim~e^{-i\bar{E} r_*}~~,~~r_*=G\ell\left[x-\log\left(\frac{x+1}{x-1}\right)\right]\,.
\end{eqnarray}
Here $r_*$ is the ``tortoise'' coordinate  \cite{Eddington,Regge:1957td,Finkelstein}, which satisfies $\lim_{r\rightarrow G\ell} r_*=-\infty$. This solution is conventionally denoted by
 $F(x)=S^{(1);\bar{\mu}}_{\bar{\nu}}(\bar{\gamma},x)$. The parameter $\bar{\nu}$ is the index of the radial spheroidal function, and is related to $\bar{\lambda},\,\bar{\mu}$ and $\bar{\gamma}$ by a transcendental equation, which is explicitly given in Appendix~\ref{app:spherser}. We explicitly checked that the other solution to the radial equation, $S^{(2);\bar{\mu}}_{\bar{\nu}}(\bar{\gamma},x)$, only leads to quantum corrections that die off in the $\hbar\rightarrow 0$ limit. The asymptotic behavior of the solution at $x\rightarrow\infty$ is given by
\begin{eqnarray}\label{eq:asymp}
S^{(1);\bar{\mu}}_{\bar{\nu}}(\bar{\gamma},x)|_{x\rightarrow\infty}\,\sim\,j_{\bar{\nu}}(\bar{\gamma} x)=j_{\bar{\nu}}(\bar{k} r)\,.
\end{eqnarray}
In appendix~\ref{app:spherser} we show how to calculate $\bar{\nu}$ explicitly. By a similar argument to the monopole case, we have $-2\delta_{j}=\pi(\nu+\hbar)$ and so the classical scattering angle is given by
\begin{eqnarray}\label{eq:thetasolNUT}
\cos\left(\frac{\chi}{2}\right)\,=\,\sqrt{1-\frac{q^2}{j^2}}\,\cos\left(\frac{\Delta\varphi}{2}\right)\,,
\end{eqnarray}
where
\begin{eqnarray}\label{eq:NUTphi}
\Delta\varphi\,=\,\pi\,\lim_{\hbar\rightarrow 0}\frac{d\nu}{d j}\,,
\end{eqnarray}
and $\nu=\lim_{\hbar\rightarrow 0}\hbar\bar{\nu}$ is given to all orders in appendix~\ref{app:spherser}. This expression coincides to all orders with the one obtained by the WKB-Hamilton-Jacobi method, equation \eqref{eq:intNUT}.

Note that although the consistency of \eqref{eq:NUTphi} and \eqref{eq:intNUT} is guaranteed by the correspondence principle between quantum and classical physics at $\hbar\rightarrow 0$, the actual all-order equality involves a highly non-trivial number theoretical identity:
\begin{eqnarray}\label{eq:number}
\pi\,\frac{d\nu}{d j}=\pi\lim_{\hbar\rightarrow0}\frac{2j}{\hbar}\frac{d\bar{\nu}}{d \bar{\lambda}}=\frac{2j}{kr_+}~K\left(\frac{r_-}{r_+}\right)\,,
\end{eqnarray}
or in other words
\begin{eqnarray}\label{eq:number2}
\lim_{s\rightarrow\infty}\,s^{-2}~\frac{d \lambda\left(s\nu,s\gamma,s\mu\right)}{d\nu}=\pi k r_+~\left[K\left(\frac{r_-}{r_+}\right)\right]^{-1}\,,
\end{eqnarray}
where $\gamma=\frac{kq}{2E},\,\mu=iq$ and $s$ is $\hbar^{-1}$. Here $\nu$ is given implicitly as the value which solves
\begin{eqnarray}
\lambda\left(s\nu,s\gamma,s\mu\right)=s^2\left[j^2-\frac{q^2}{2}\left(\beta^2+3\right)\right]-\frac{1}{4}\,,
\end{eqnarray}
where $\beta=k/E$. The function $\lambda(\nu,\gamma,\mu)$ is known in the literature as the (analytically continued) spheroidal eigenvalue \cite{meixner2006mathieu,Falloon_2003}. Here we uncover a very non-trivial relation between its derivative in the limit $\nu,\gamma,\mu\rightarrow\infty$ and the elliptic integral $K$. We are not aware of previous derivations of this relation in the literature.

\section{Outlook: Towards Non-Perturbative Self-Force Calculations}\label{sec:outlook}

Our results clearly demonstrate that classical effects that appear to be non-perturbative in the PM expansion can be fully captured in the $\hbar\rightarrow 0$ limit of quantum wave equations. In and of itself, this should not surprise the reader much, as it is a consequence of the correspondence principle between quantum and classical physics.  
However, we uncovered in detail exactly \textit{how} the quantum amplitude encodes the classical scattering data in the $\hbar\rightarrow 0$ limit, namely
\begin{itemize}
    \item The phase shifts go over to their WKB-Hamilton-Jacobi values, which are in turn related to the classical radial action by \eqref{eq:relation}.
    \item The Poisson resummed partial wave decomposition has a saddle point at the classical $j$, which in turn leads to the classical scattering angle \eqref{eq:thetasol} (cf. \eqref{eq:poirelch} for the scalar case).
\end{itemize}
We applied this formalism in the probe limit to calculate classical effects that are inherently non-perturbative from the qunatum point of view. For example, we reproduced the winding of classical trajectories for relativistic Coulomb and Schwarzschild scattering, as well as the effect of the extra angular momentum $-q\hat{r}$ in the gravitational (electromagnetic) field for probe mass-NUT (charge-monopole) scattering. In the two latter cases, we also correctly reproduced the fact that the classical trajectories are confined to a cone around the total angular momentum $\vec{J}$.

Finally, our quantum-classical matching uncovers previously unknown (to us) number-theoretic relations such as \eqref{eq:number2}. We expect a similar number-theoretic relation to hold in the context of probe scattering off Schwarzschild: that is, between the phase shift emerging from the $\text{HeunC}$ function, and the Appell F1 function which we know describes the classical trajectory.

An obvious direction for future work is to apply similar methods to black hole scattering away from the probe limit, i.e. to all orders in $G/j$ but perturbatively in $\kappa=\frac{m_1m_2}{(m_1+m_2)^2}$. In other words, it would be interesting to apply our method to calculate $\mathcal{O}(\kappa^n)$ ``self-force'' corrections. 
Since our method is non-perturbative in $G/j$, its application away from the probe limit will inevitably involve both energy loss to radiation and conservative tail effects that are nonlocal in time \cite{BonnorRotenberg66,BlanchetDamour86,BlanchetDamourTail,BlanchetDamour92,Blanchet_1993,Blanchet:1997jj,Asada:1997zu,Galley:2015kus,Marchand:2016vox}.

One possible way forward would be to consider quantum scattering in the full Arnowitt-Deser-Misner Hamiltonian \cite{ADM,Schafer:2018kuf}, \textit{without} integrating out the gravitational field. This would lead to a set of coupled wave equations for the two black holes and the gravitational field, which could be solved to all orders in $G/j$ but order-by-order in $\kappa$. To focus on conservative dynamics, we could impose the boundary condition of a pure Schwarzschild metric at $r\rightarrow\infty$, such that there is no leakage of energy via gravitational waves. However, tail effect will be captured since outgoing gravitational waves would be reflected back to the center by the ambient Schwarzschild metric in the far zone.

Calculating self-force corrections in terms of wave equations would have the additional advantage of smoothing out the inherent divergences which are ubiquitous in coupling point masses to GR, which requires very careful regularization in the standard treatments \cite{Barack:1999wf,Blanchet:2000cw,Damour:2001bu,Blanchet:2013haa,Schafer:2018kuf,Barack:2018yly,Barack:2018yvs}.

Turning away from gravitational wave physics, our methods may have an application to the study of the double copy beyond perturbation theory~\cite{Monteiro:2014cda,Luna:2015paa,Adamo:2017nia,Adamo:2018mpq,Monteiro:2020plf,Campiglia:2021srh,Borsten:2021hua,Chacon:2021wbr,Gonzo:2021drq,Godazgar:2021iae,Adamo:2021dfg}. It is by now well-established that a pure NUT is related to the magnetic monopole by the double copy in perturbation theory; indeed, this was an initial motivation for our work on the pure NUT. We have now seen how to construct amplitudes for monopoles and NUTs to all orders, so we have the theoretical data to explore this double copy to all orders. Of course this would just be a prelude to the study of the double copy relating a charge and Schwarzschild to all orders.

\section*{Acknowledgements}

We thank Yu-tin Huang for many helpful discussions, Mao Zeng for insightful and encouraging correspondence, and Andr\'es Luna for reading our manuscript and pointing out a clerical error in a key formula. We thank the Galileo Galilei Institute for Theoretical Physics (GGI) for hosting a workshop, conference and training week on ``Gravitational scattering, inspiral, and radiation'' which informed and enriched our work.
DOC is supported by the STFC grant ST/P0000630/1. OT is supported in part by the DOE under grant DE-AC02-05CH11231.

\appendix
\section{Characteristic Equation for the NUT Spheroidal Equation}\label{app:spherser}
In this appendix we calculate the index $\bar{\nu}$ for the prolate spheroidal equation Eq.~\ref{eq:radTaubsph}. The index is linked to the scattering phase shift by $-2\delta_{j}=\pi\,\lim_{\hbar\rightarrow 0}\nu$. The index $\bar{\nu}$ is related to the parameters of the spheroidal equation $\bar{\lambda},\,\bar{\mu},\,\bar{\gamma}$ by a transcendetal equation. Following \cite{Falloon_2003}, we define the following variables:
\begin{eqnarray}\label{eq:recurco}
A_{2k}\,&=&\,-\bar{\gamma}^2\frac{(\bar{\nu}-\bar{\mu}+2k-1)(\bar{\nu}-\bar{\mu}+2k)}{(2\bar{\nu}+4k-3)(2\bar{\nu}+4k-1)}\nonumber\\[5pt]
B_{2k}\,&=&\,\left(\bar{\nu}+2k\right)\left(\bar{\nu}+2k+1\right)\,-\,2\bar{\gamma}^2\,\frac{\left(\bar{\nu}+2k\right)\left(\bar{\nu}+2k+1\right)+\bar{\mu}^2-1}{(2\bar{\nu}+4k+3)(2\bar{\nu}+4k-1)}\nonumber\\[5pt]
C_{2k}\,&=&\,-\bar{\gamma}^2\frac{(\bar{\nu}+\bar{\mu}+2k+1)(\bar{\nu}+\bar{\mu}+2k+2)}{(2\bar{\nu}+4k+3)(2\bar{\nu}+4k+5)}\,,
\end{eqnarray}
as well as
\begin{eqnarray}\label{eq:recurco2}
\alpha_{2k}\,&=&\,A_{2k}C_{2k-2}\nonumber\\[5pt]
\beta_{2k}\,&=&\,B_{2k}\,.
\end{eqnarray}
The variables $\bar{\lambda},\,\bar{\gamma}$ and $\bar{\mu}$ for NUT scattering are defined in \eqref{eq:lamb}. In this appendix $k$ is an integer index, not to be confused with the plane wave momentum $k$.
In terms of $\alpha_{2k}$, $\beta_{2k}$, we define two continued fractions, whose sum is required to vanish. These are
\begin{eqnarray}\label{eq:contf}
F_+\,&=&\,\beta_0-\bar{\lambda}-\frac{\alpha_{0}}{\beta_{-2}-\bar{\lambda}-}\frac{\alpha_{-2}}{\beta_{-4}-\bar{\lambda}-}\ldots\nonumber\\[5pt]
F_-\,&=&\,-\frac{\alpha_{2}}{\beta_{2}-\bar{\lambda}-}\frac{\alpha_{4}}{\beta_{4}-\bar{\lambda}-}\ldots
\end{eqnarray}
The transcendental equation for $\nu$ is then given by 
\begin{eqnarray}\label{eq:eqnu}
F_+~+~F_-\,=\,0\,.
\end{eqnarray}
Note that it is usually treated as an equation for $\bar{\lambda}$, which is known as the ``spheroidal eigenvalue'' of the problem. In this case the index $\bar{\nu}$ is treated as an integer enumerating the eigenvalue $\bar{\lambda}(\bar{\nu})$. In particular, $\bar{\nu}$ and $\bar{\lambda}(\bar{\nu})$ related by Eq.~\ref{eq:eqnu} satisfy the continuity relation $\lim\limits_{\bar{\gamma}\rightarrow 0}\bar{\lambda}(\bar{\nu})=\bar{\nu}(\bar{\nu}+1)$. Here we use the same machinery in a different manner, by fixing $\bar{\lambda},\,\bar{\gamma},\,\bar{\mu}$ and solving for $\bar{\nu}$. It is particularly useful for our purposes to substitute an ansatz for $\bar{\nu}$ of the form
\begin{eqnarray}\label{eq:nuansz}
\bar{\nu}\,=\,\sqrt{\bar{\lambda}+\sum_{i=0}^\infty\,c_i\bar{\lambda}^{-i}}-\frac{1}{2}\,,
\end{eqnarray}
We can solve for the coefficients explicitly by substituing this ansatz in the transcendental relation \eqref{eq:eqnu}. The first 4 coefficients (corresponding to a 6PM expansion) are then 
\begin{eqnarray}\label{eq:nucos}
&&c_0\,=\,\frac{1}{4}\,\left(2\bar{\gamma}^2+1\right)\nonumber\\[5pt]
&&c_1\,=\,-\frac{\bar{\gamma}^2}{32}\,\left[\bar{\gamma}^2-4\,(4\bar{\mu}^2-1)\right]\nonumber\\[5pt]
&&c_2\,=\,\frac{\bar{\gamma}^2}{128}\,\left[2\bar{\gamma}^4-(8\bar{\mu}^2+1)\,\bar{\gamma}^2+12\,(4\bar{\mu}^2-1)\right]\nonumber\\[5pt]
&&c_3\,=\,-\frac{\bar{\gamma}^2}{8192}\,\left[77\bar{\gamma}^6-96\,(4\bar{\mu}^2+1)\,\bar{\gamma}^4+32\,(104\bar{\mu}^4-172\bar{\mu}^2+41)\,\bar{\gamma}^2-576\,(4\bar{\mu}^2-1)\right]\,.
\end{eqnarray}
See \eqref{eq:lamb} for the definitions of $\bar{\lambda},\,\bar{\gamma},\,\bar{\mu}$.

\bibliographystyle{JHEP}
\providecommand{\href}[2]{#2}\begingroup\raggedright\endgroup

\end{document}